\documentclass[a4paper, amsfonts, amssymb, amsmath, reprint, showkeys, nofootinbib, twoside, superscriptaddress]{revtex4-1}
\usepackage[english]{babel}
\usepackage[utf8]{inputenc}
\usepackage[colorinlistoftodos, color=green!40, prependcaption]{todonotes}

\usepackage[pdftex, pdftitle={Article}, pdfauthor={Author}]{hyperref} % For hyperlinks in the PDF

\usepackage{graphicx}% Include figure files
\usepackage{dcolumn}% Align table columns on decimal point
\usepackage{bm}% bold math
\usepackage{hyperref}% add hypertext capabilities
\usepackage[mathlines]{lineno}% Enable numbering of text and display math
\usepackage{siunitx}%Nice number and unit formatting
\usepackage{xspace}%spaces at end of custom-defined macros
\usepackage{epstopdf}%Import eps files from MATLAB- Has to be included AFTER graphicx
\usepackage[]{pgfplots}
\pgfplotsset{compat=1.12}
\usetikzlibrary{arrows}
\usepackage[final]{changes}

\usepackage[eulergreek]{sansmath}

\usepackage[]{natbib}

\usepackage[caption=false]{subfig}

\begin{document}
\author{Ehsan Hassanpour}
    \thanks{These two authors contributed equally}
    %\email[Correspondence email address:~]{ehsan.hassanpour@mat.ethz.ch}% Your name
    \affiliation{Department of Materials, ETH Zurich, Vladimir-Prelog-Weg 4, 8093 Zurich, Switzerland}
    \author{Mads~C.~Weber}
    \thanks{These two authors contributed equally}
    \affiliation{Department of Materials, ETH Zurich, Vladimir-Prelog-Weg 4, 8093 Zurich, Switzerland}
     \author{Amad\'{e} Bortis}
     \affiliation{Department of Materials, ETH Zurich, Vladimir-Prelog-Weg 4, 8093 Zurich, Switzerland}
     \author{Yusuke Tokunaga}
     \affiliation{University of Tokyo, Department of Advanced Materials Science, Kashiwa 277-8561, Japan}
     \author{Yasujiro Taguchi}
     \affiliation{RIKEN Center for Emergent Matter Science (CEMS), Wako 351-0198, Japan}
     \author{Yoshinori Tokura}
     \affiliation{RIKEN Center for Emergent Matter Science (CEMS), Wako 351-0198, Japan}
     \affiliation{Department of Applied Physics, University of Tokyo, Tokyo 113-8656, Japan}
     \author{Andres Cano}
     \affiliation{Institut N\'eel, CNRS \& Universit\'e Grenoble Alpes, 38042 Grenoble, France}
     \author{Thomas Lottermoser}
     \affiliation{Department of Materials, ETH Zurich, Vladimir-Prelog-Weg 4, 8093 Zurich, Switzerland}     
     \author{Manfred Fiebig}
     \email[Correspondence email address:~]{manfred.fiebig@mat.ethz.ch}
     \affiliation{Department of Materials, ETH Zurich, Vladimir-Prelog-Weg 4, 8093 Zurich, Switzerland}

\title{Interconversion of multiferroic domains and domain walls}

\date{\today}
\begin{abstract}
Materials with long-range order like ferromagnetism or ferroelectricity exhibit uniform, yet differently oriented three-dimensional regions called domains that are separated by two-dimensional topological defects termed domain walls\cite{Tagantsev2010,AlexHubert1998}. A change of the ordered state across a domain wall can lead to local non-bulk properties such as enhanced conductance or the promotion of unusual phases\cite{Seidel2009,Meier2012,Farokhipoor2014}.  Although highly desirable, controlled transfer of these exciting properties between the bulk and the walls is usually not possible. Here we demonstrate this crossover from three- to two-dimensions for confining multiferroic Dy$_{0.7}$Tb$_{0.3}$FeO$_3$ domains into multiferroic domain walls at a specified location within a non-multiferroic environment. This process is fully reversible; an applied magnetic or electric field controls the transformation. Aside from the aspect of magnetoelectric functionality, such interconversion can be key to tailoring elusive domain architectures such as in antiferromagnets.
\end{abstract}0


\maketitle
%\section{Introduction}
Recently, the interests in ferroic materials with magnetic or electric order evolved from the three-dimensional (3D) bulk domains to the two-dimensional (2D) domain walls. As inherent inhomogeneity, domain walls are a source of specific phenomena that are forbidden in the uniform interior of the corresponding domains. Examples are the occurrence of magnetization, polarization, magnetoelectric coupling, (super-)conductivity, memory effects or a change of melting temperature\cite{Farokhipoor2014,VanAert2012,Aird1998,Fiebig2002,Seidel2009,Meier2012,Schroder2012,Beilsten-Edmands2016,Salje2010,Catalan2012,Meier2015,Logginov2007}. Hence, domain walls may be regarded as novel state of a material, virtually a world in its own self and seemingly separated from the surrounding bulk phase.

It would now be very desirable if the dimensional limitation of these exciting phenomena could be overcome. For example, one could consider the confinement of a multifunctional bulk state into domain walls where they could establish a form of rewritable electromagnetic circuits\cite{Crassous2011,Aird1998}. Reversely, the domain walls may seed the recovery of the original bulk state, acting as its memory\cite{Taniguchi2009,Beilsten-Edmands2016}.

A class of materials in which the transition from the 3D bulk to the 2D domain walls could be virtually continuous are compounds with phase transitions in which the symmetry of the domains and domain walls coincide crosswise on either side of the phase boundary. In such materials, domain walls can in principle gradually transform into domains, and vice versa, across a first-order phase transition. This concept was discussed theoretically\cite{Mitsek1970,Belov1976,Baryakhtar1988} and systems showing a behaviour consistent with this concept were reported\cite{Gnatchenko1981,Kharchenko1989}. Conclusive observation of a smooth, deterministic transfer of a domain wall into a domain, however, and, in particular, the reversible interconversion between domains and domain walls have not been presented. Notably, the practical consequences of such interconversion for the functionalities of materials were never debated.

We demonstrate this interconvertibility and apply it to tune a multiferroic state between three and two dimensions in a fully controlled way. Specifically, we confine a multiferroic bulk state into a multiferroic domain wall at a specified location within a non-multiferroic environment. We act on this 2D magnetoelectric object with magnetic or electric fields and evidence the presence of a switchable magnetization and polarization in the wall. We furthermore employ the fields to transfer the multiferroic domain wall back into a multiferroic bulk state. We then discuss the general occurrence and benefits of an ordered state with controllable transfer in between 3D and 2D.

\begin{figure*}[h]
	\centering
	\includegraphics[width=0.75\textwidth]{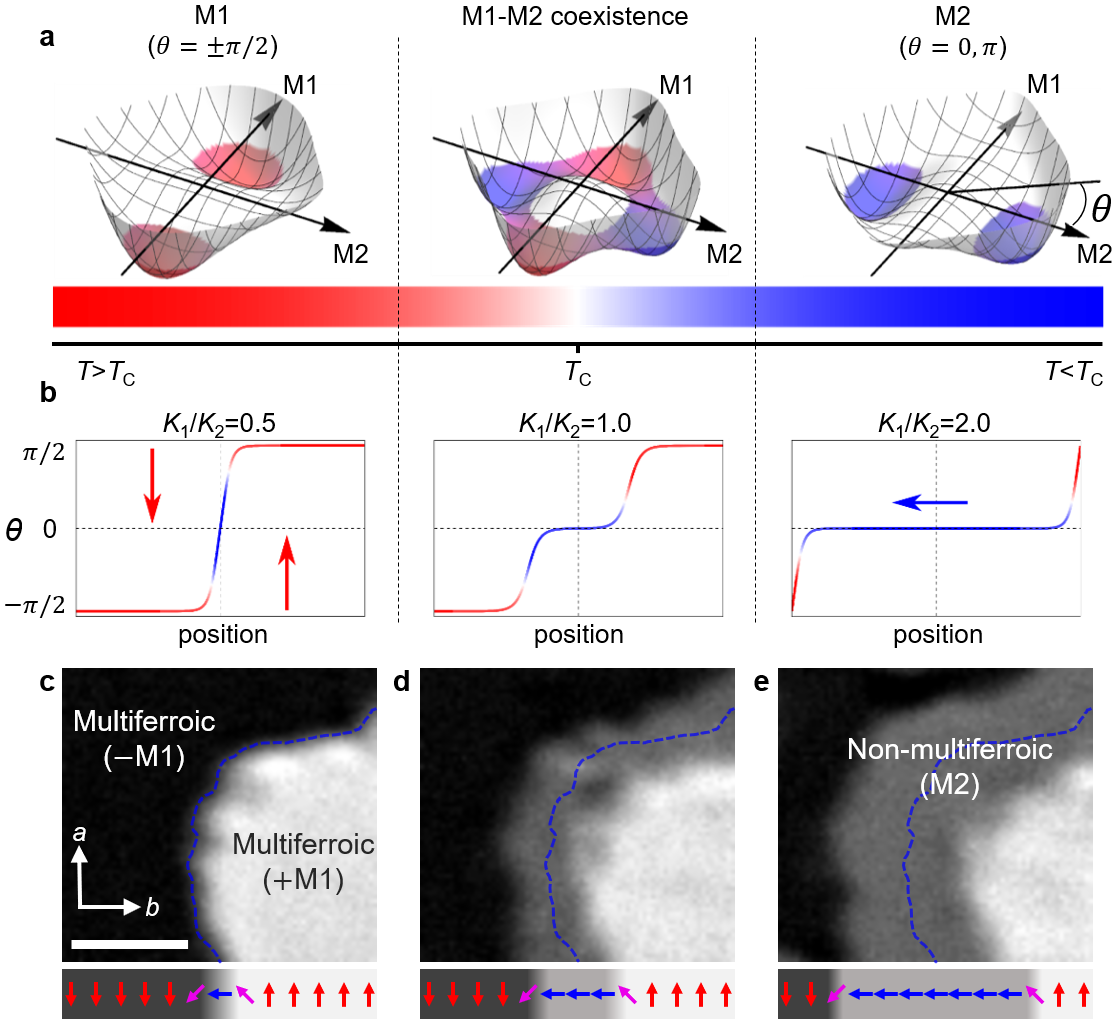}
	\caption{%
		\label{model}
		\textbf{Interconversion of domains and domain walls.} {\bf a}, Temperature dependence of the free-energy landscape for the first-order phase transition of Eq.~\ref{hamil} with $\theta$ as orientation of the order parameter. {\bf b}, Evolution of domains and domain walls across the $M1\leftrightarrow M2$ phase transition. Domain walls expand into domains and domains are confined into domain walls, with an intermediate regime where the $M1$ and $M2$ phases coexist. $K_1$ and $K_2$ are magnetocrystalline anisotropy parameters. Blue and red arrows depict the orientation $\theta$ of the order parameter.
		\textbf{c}-\textbf{e}, Upon temperature decrease, a domain wall (dashed blue line) separating opposite multiferroic domains in c expands and transforms to a domain of the antiferromagnetic, non-multiferroic phase in c and d. Faraday rotation (FR) by $\phi_{\rm FR}$ shows the $M1$ domains as dark ($\phi_{\rm FR}<0$) and bright ($\phi_{\rm FR}>0$) regions and the $M2$ domains ($\phi_{\rm FR}=0$) as grey area. Scale bar, 100{\,}$\mu$m.
	}
\end{figure*}

%\section{Theory}
For the reasons given below, we choose the rare-earth orthoferrite Dy$_{0.7}$Tb$_{0.3}$FeO$_3$ (see Ref.~\onlinecite{Tokunaga2012} about sample preparation) as our model system. Multiferroicity occurs at $T_C \simeq 2.65$~K, caused by simultaneous antiferromagnetic ordering of the rare-earth and iron spin systems, coupled by exchange striction\cite{Tokunaga2012}. An electric polarization $P_s=0.12$~$\mu$C/cm$^2$ coexists with a Dzyaloshinskii-Moriya-type weak ferromagnetic moment $M_s=0.15$~$\mu_B$ per formula unit, both oriented along the $c$-axis. Subsequently, at 2.3~K a first-order spin reorientation of the iron sublattice from the multiferroic to antiferromagnetic phase with $M_s$, $P_s=0$ occurs. The simplest possible form to describe such a transition in between two magnetic phases M1 and M2 is the phenomenological effective Hamiltonian\cite{Belov1976,Baryakhtar1988}
\begin{equation} \label{hamil}
H = A |\nabla \theta|^2 + K_1 \sin^2 \theta  + K_2 \sin^4 \theta.
\end{equation}
As depicted in Fig.~\ref{model}, the angle $\theta$ distinguishes the domain states on either side of the $M1 \leftrightarrow M2$ phase transition. $K_1$ and $K_2$ are magnetocrystalline anisotropy parameters. $K_1 = 0$ and $K_1 = - 2K_2$ set the limits of the coexistence region. The gradient term with $A$ as exchange constant further determines the order-parameter rotation across the domain walls. 

In Dy$_{0.7}$Tb$_{0.3}$FeO$_3$, $M1$ and $M2$ are associated with the multiferroic (weak ferromagnetic) and non-multiferroic (purely antiferromagnetic) phases, respectively, with $\theta_{\pm M1} = +\pi/2, -\pi/2$ and $\theta_{\pm M2} = 0, \pi$. Figure~\ref{model}b depicts the evolution according to Eq.~(\ref{hamil}) of a configuration starting with two well-defined domains outside the coexistence region separated by a single domain-wall\cite{Mitsek1970}. By entering the $M1/M2$ coexistence region, the domain wall widens and transforms into a domain of the other phase. Conversely, the initial domain shrinks and eventually transforms into a domain wall separating the new domains on the opposite side of the coexistence region. Hence, the first-order phase transition between the multiferroic $M1$ and the antiferromagnetic $M2$ phase in Dy$_{0.7}$Tb$_{0.3}$FeO$_3$ should in principle allow us to reversibly convert a multiferroic domain into a multiferroic domain wall in an antiferromagnetic, non-multiferroic environment.

%\section{Results}

We first verify the theoretically predicted transformation of a domain wall into a domain\cite{Mitsek1970} experimentally. In Dy$_{0.7}$Tb$_{0.3}$FeO$_3$, we can exploit the magnetization of the multiferroic phase to distinguish the $\pm M1$ and $M2$ domain states by spatially resolved real-time imaging experiments, using the magneto-optical Faraday effect as magnetization probe. Furthermore, Dy$_{0.7}$Tb$_{0.3}$FeO$_3$ exhibits a relatively broad region of phase coexistence. This gives us ample time to image the phase coexistence while tuning the balance between the competing multiferroic and antiferromagnetic phases.

Figure~\ref{model}c-e shows sequential Faraday images of a $c$-oriented Dy$_{0.7}$Tb$_{0.3}$FeO$_3$ sample cooled across the multiferroic-to-antiferromagnetic transition. The corresponding order-parameter distribution is sketched beneath each image with arrows representing the local value of $\theta$ in Eq.~(\ref{hamil}). We refrain from quantifying $K_1$ and $K_2$ since their absolute values are irrelevant for the further course of this manuscript. In Fig.~\ref{model}c we see a $\pm M1$ domain pair at $T>T_C$. Because of the opposite direction of magnetization and, hence, Faraday rotation, the domains appear as bright and dark regions. Figures~\ref{model}d and \ref{model}e show the same region about 4 and 8 minutes after cooling the sample to $T\lesssim T_C$. A grey stripe centers at the domain-wall position of Fig.~\ref{model}c that widens with time. Its zero Faraday rotation identifies it as antiferromagnetic region. Below $T_C$, this is the dominating phase and, starting from the original $\pm M1$ domain wall, we perceive the homogeneous expansion of this wall into a $M2$ domain engulfing the sample as time progresses. Note that this expansion of the antiferromagnetic phase occurs in a deterministic way, starting uniformly from the center of the domain wall of the multiferroic phase. This is in stark contrast to nucleation on random structural defects as in most first-order phase transitions. It is also in contrast to random, needle-like emergence of domains of the target phase at or within domain walls of the seed phase, or to a coexistence of phases in stripe domains of variable width.\cite{King1973}

\begin{figure*}
	\centering
	\includegraphics[width=0.75\textwidth]{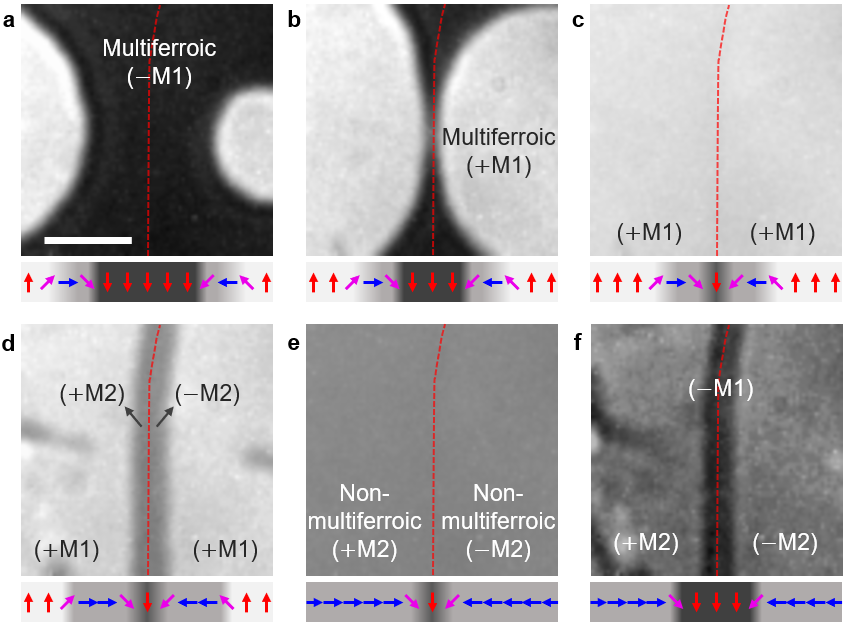}
	\caption{%
		\label{d-to-w} 
		\textbf{Transformation of a multiferroic bulk domain into a multiferroic domain wall in a non-multiferroic environment.} Panels a-f show the same area; scale bar, 100{\,}$\mu$m. Beneath each image, the magnetic structure is sketched with arrows indicating the order-parameter orientation ($\theta$ in Eq.~\ref{hamil} and Fig.~\ref{model}a).
		\textbf{a}-\textbf{c}, Starting from a single-domain state ($-M1$, dark), two opposite multiferroic domains ($+M1$, bright) are generated by application of a magnetic field $\mu_0H_c\leq 100${\,}mT. The $+M1$ domains grow until they meet along the red dashed line. Here, the two domain walls in a and b have merged into a single topological object.
		\textbf{d}, \textbf{e}, Setting $T=1.8$~${\rm K}<T_c$ at $\mu_0H_c=0$ returns the sample to the antiferromagnetic $M2$ state. The wall separating the $+M1$ domains in c expands into an $M2$ domain (grey) gradually engulfing the entire visible area. Note that e is composed of opposite $M2$ domains separated by a multiferroic domain wall that we cannot visualize because of the limited resolution of our experiment.
		\textbf{f}, However, heating the sample to 2.3{\,}K stabilizes the multiferroic phase and lets the suspected multiferroic domain wall in e expand into a $-M1$ domain, thus confirming its presence.
	}
\end{figure*}

Now we employ this deterministic aspect for confining a multiferroic domain into a multiferroic domain wall at a specified location within a non-multiferroic environment. We begin at $T>T_C$ with a $-M1$ domain sandwiched between $+M1$ domains (Fig.~\ref{d-to-w}a). In a magnetic field we enlarge the $+M1$ domains (Fig.~\ref{d-to-w}b) until they meet so that the $-M1$ domain is no longer detectable (Fig.~\ref{d-to-w}c). After cooling the sample to $T\lesssim T_C$, the antiferromagnetic state emerges (Fig.~\ref{d-to-w}d) and grows with time until it has filled the entire field of view (Fig.~\ref{d-to-w}e). Note that the $M2$ state originates from the meeting position of the $+M1$ domains in Fig.~\ref{d-to-w}c, evidencing the presence of a remaining topological object at this location that seeds the $M2$ order.

\begin{figure}
	\centering
	\includegraphics[width=0.4\textwidth]{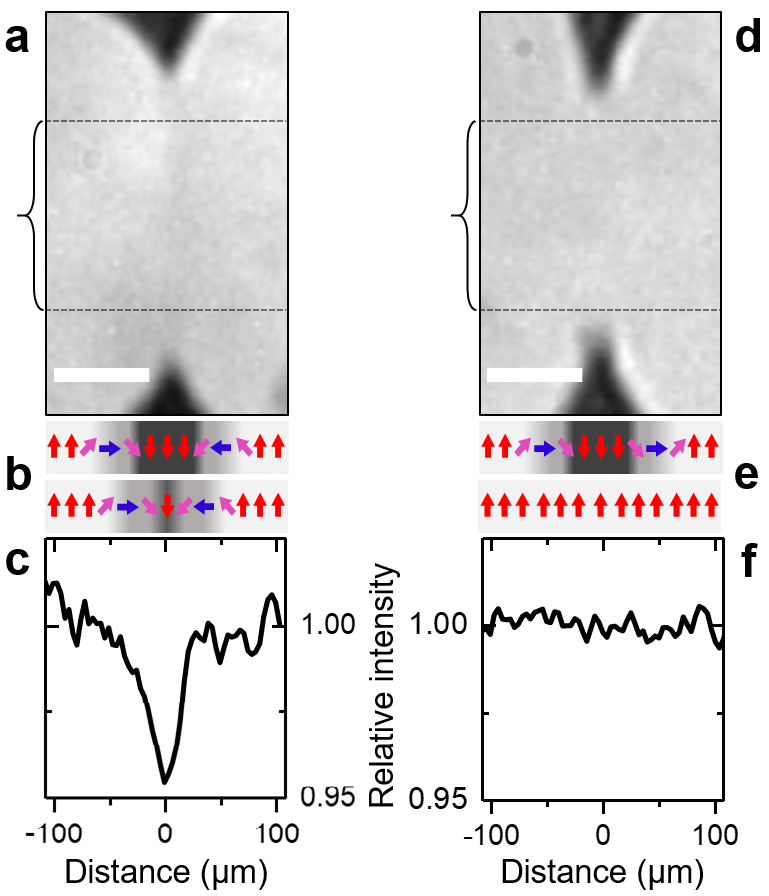}
	\caption{%
		\label{skyrmion}
		\textbf{Non-coalescing and coalescing multiferroic domains.}
		\textbf{a}, Meeting of two multiferroic $+M1$ domains obtained following the procedure as in Fig.~\ref{d-to-w}a-c. Scale bars, 100{\,}$\mu$m. Scenario with same sense of rotation of the order parameter across the $+M1$ $\to$ $-M1$ domain wall for both $+M1$ domains.
		\textbf{b}, Order-parameter orientation (arrows) across the boundary of the $+M1$ walls before (top) and after (bottom) the meeting of the $+M1$ domains. 
		\textbf{c}, Horizontal intensity scan of the image in {\bf a}, vertically averaged across the area indicated by the curly bracket.
		\textbf{d}-\textbf{f}, Same as a-c, but with opposite sense of rotation of the order parameter across the $+M1$ $\to$ $-M1$ domain wall for the two $+M1$ domains. Note that the $+M1$ domains coalesce for opposite sense of rotation of the order parameter in the meeting domains. The $+M1$ domains do not coalesce if the sense of rotation is the same. Instead, a  $360^{\circ}$ domain wall is formed with the identity of a one-dimensional N\'{e}el-like antiferromagnetic skyrmion. At the position of this skyrmion the image brightness exhibits a dip caused by the local variation of the Faraday rotation across the wall.
	}
\end{figure}

This topological object is scrutinized in Figure~\ref{skyrmion}. As we see, there are two types of meetings between $+M1$ domains on a $-M1$ background. They reflect the two types of domain walls a multiferroic domain is expected to exhibit, namely $+M1$ $\to$ $-M1$ walls with either clockwise or counterclockwise rotation of spins across the wall.\cite{Houchmandzadeh1991} When domain walls of the opposite type meet, the respective spin rotations cancel and the walls annihilate so that the meeting domains coalesce. Alternatively, for a meeting of the same type of domain walls, a $360^{\circ}$ spin rotation as sketched in Fig.~\ref{skyrmion}b occurs. This object, equivalent to a topologically protected antiferromagnetic 1D-skyrmion\cite{Bergmann2014}, prevents the coalescence of the domains as indicated by the brightness dip in Fig.~\ref{skyrmion}c. Instead, it can seed a $+M2$ and a $-M2$ domain as sketched beneath Fig.~\ref{d-to-w}d,e. Note that a magnetic field three times the value of the coercive field is required to destroy the $360^{\circ}$ wall.

Hence, Fig.~\ref{d-to-w}e shows a $+M2$ and a $-M2$ domain; according to Fig.~\ref{model}a, they are separated by a multiferroic domain wall into which the multiferroic bulk state of Fig.~\ref{d-to-w}a has been confined. This wall has been placed at the meeting point of the two former $+M1$ domains in Figs.~\ref{d-to-w}c. For confirmation, we re-heat the sample to $T\gtrsim T_C$; Fig.~\ref{d-to-w}f shows that this reconverts the suspected multiferroic domain wall into a multiferroic $-M1$ domain. 

\begin{figure}
	\centering
	\includegraphics[width=0.40\textwidth]{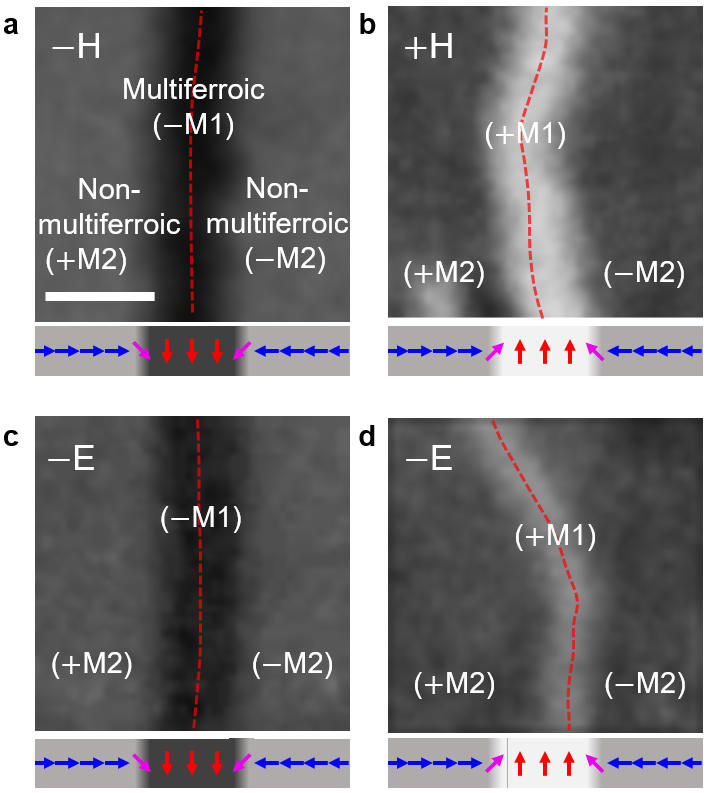}
	\caption{%
		\label{MF-wall}
		\textbf{Magnetoelectric field control of multiferroic domain walls in a non-multiferroic environment.} 
		Panels a to d show equally sized regions on the same sample, yet imaged on different areas and in separate experiments. Scale bar, 30{\,}$\mu$m. For experiments in a-c, a state as in Fig.~\ref{d-to-w}e with $+M2$ and $-M2$ domains separated by a $-M1$-type domain wall was prepared while for d, opposite $M2$ domains are separated by a $+M1$-type domain wall. Magnetic and electric fields are applied as labeled, with the $+$ sign indicating out-of-plane orientation. Beneath each image, the magnetic structure is sketched with arrows indicating the order-parameter orientation. \textbf{a},\textbf{b}   A magnetic field $\pm H$ sets the magnetization of the multiferroic wall to $\pm M_s$ and expands it into a $\pm M1$ multiferroic bulk domain with a 1:1 correspondence between the signs of \textit{H} and $M_s$. \textbf{c},\textbf{d}   An electric field $\pm E$ sets the polarization of the multiferroic wall to $\pm P_s$ and expands it into a multiferroic $M1$ bulk domain. Because of the involvement of three order parameters in the coupling, the sign of $M_s$ is not determined by the sign of $P_s(E)$ but rather by the history of the sample\cite{Tokunaga2009}. This is highlighted by c and d, where a field $-E$ induces a $+M1$ and $-M1$ domain, respectively.
	}
\end{figure}

To describe the wall as multiferroic, however, it is not sufficient to show that we can convert it into a multiferroic domain, but rather to demonstrate the coexistence of a polarization and a magnetization for the wall itself. This is done in Fig.~\ref{MF-wall} which shows the response of domain walls generated as in Fig.~\ref{d-to-w}e to static $c$-oriented magnetic or electric fields. We see that either field initiates the transfer of the wall back into a  domain of the multiferroic phase, even though temperature-wise the material still favours the antiferromagnetic phase. This is only possible, if the wall already has a magnetization $M_s$ and a polarization $P_s$ the respective applied field can act on to initiate the transfer. Furthermore, the transfer is energetically beneficial only if $M_s$ or $P_s$ points in the direction of the applied field. Since the field has triggered the transfer in all our experiments the wall magnetization and polarization itself must be switchable and hence set the direction of $M_s$ or $P_s$ of the expanding domain. Determination of the sign of $M_s$ by the magnetic field is directly seen in Figs.~\ref{MF-wall}a,b, whereas Fig.~\ref{MF-wall}c shows that the electric field acts on the magnetization via magnetoelectric coupling. (Note that because of the involvement of three order parameters in this coupling, the sign of $M_s$ is not determined by the sign of $P_s(E)$ but rather by the history of the sample\cite{Tokunaga2009}, as Fig.~\ref{MF-wall}d shows.) We thus see that despite the confinement and the non-multiferroic environment, a wall as in Fig.~\ref{d-to-w}e retains the properties of the multiferroic bulk phase.

%\section{Conclusion}
To conclude, we have experimentally demonstrated the interconversion of a multiferroic state in between three and two dimensions. We have confined the multiferroic bulk phase into a multiferroic domain wall at a predetermined location inside a non-multiferroic environment. The initial multiferroic hallmark properties, namely a magnetization and a polarization that are coupled and switched by an applied field, are retained by the wall. Magnetic or electric fields act on the wall and convert it back into a bulk multiferroic domain with a predetermined direction of magnetization or polarization.

The concept of interconversion can be expanded beyond multiferroics. First, there are plenty of first-order magnetic phase transitions fulfilling the rather basic conditions of Eq.~(\ref{hamil}) with CaFe$_2$O$_4$ as a very recent example to catch scientific attention\cite{Stock2017}. Note that even if the region of phase coexistence may be narrower than in our case, possibly even too narrow for convenient experimental probing during the transition, a system will still be subject to the same dynamics as found in Dy$_{0.7}$Tb$_{0.3}$FeO$_3$. Even non-magnetic types of order can show this phenomenon. For example, a (static) expansion of domain walls into domains with phase coexistence based on a chemical gradient rather than the first-order nature of a phase transition was observed in the ferroelectric-to-nonferroelectric transition in InMnO$_3${\cite{Huang2014}}. Even organic materials may show a transition between phases with neutral and ionic building blocks emerging out of an expanding singularity by progressive charge transfer\cite{Nagaosa1986,Horiuchi2000}.

Second, our work expands the possibilities for functionalizing domain walls. Unlike in previous examples\cite{Logginov2007}, the magnetoelectric properties are not confined to the domain walls but can be expanded into the bulk when needed. The magnetoelectric properties of the walls may be employed to visualize or even tailor domain patterns in materials whose order is normally difficult to access, such as $180^{\circ}$ antiferromagnets. Finally, our work provides a rare opportunity of deterministic nucleation in a first-order phase transition since the nucleating phase is seeded by the order and symmetry in domain walls rather than occurring on random defects.

%\section*{Acknowledgement}
We thank Dr. Morgan Trassin for deposition of Pt electrodes on the samples. This work was financially supported by SNSF (Grant No. 200021\_178825/1) and European Research Council (Advanced Grant 694955-INSEETO). Y. Tokunaga was supported by JSPS Grant-in-Aids for Young Scientists (A) (No. 25707032). Y. Tokunaga, Y. Taguchi and Y. Tokura were supported by the Japan Society for the Promotion of Science (JSAP) through its ``Funding Program for World-Leading Innovative R\&D on Science and Technology'' (FIRST Program). M. F. thanks ETH Zurich and CEMS at RIKEN for support of his research sabbatical.

\bibliography{domain-dw-duality}

\end{document}